%% file: paper.tex
\pdfoutput=1

\documentclass[sigconf]{acmart}

\renewcommand\footnotetextcopyrightpermission[1]{} % removes footnote with conference information in first column
\usepackage{booktabs} % For formal tables
\usepackage{makecell}
\usepackage{xspace}
\usepackage{graphicx}
\usepackage{bmpsize}
\usepackage{color}
\definecolor{red}{rgb}{1,0,0}
\definecolor{green}{rgb}{0,1,0}
\definecolor{blue}{rgb}{0,0,1}
\definecolor{cyan}{rgb}{0.4,1,1}
\definecolor{orange}{rgb}{1,0.6,0}
\definecolor{dkgreen}{rgb}{0,0.6,0}
\definecolor{dkred}{rgb}{0.6,0,0}
\definecolor{gray}{rgb}{0.5,0.5,0.5}
\definecolor{purple}{rgb}{0.58,0,0.82}

\usepackage[font=small,labelfont=bf]{caption}

\newif\ifDRAFT
\DRAFTfalse

\input{urls}
\input{abbr}
\input{data}
\input{commands}

\begin{document}
\sloppy
\title{SpeedReader: Reader Mode Made Fast and Private}

\author{Mohammad Ghasemisharif}
\affiliation{%
  \institution{University of Illinois at Chicago}
}
\email{mghas2@uic.edu}

\author{Peter Snyder}
\affiliation{%
  \institution{Brave Software}
}
\email{pes@brave.com}

\author{Andrius Aucinas}
\affiliation{%
  \institution{Brave Software}
}
\email{aaucinas@brave.com}

\author{Benjamin Livshits}
\affiliation{%
  \institution{Brave Software / Imperial College London}
}
\email{ben@brave.com}


\input{0_abstract}

\maketitle
\pagestyle{plain}

\input{1_introduction}
\input{2_background}

\input{3_classifier}
\input{4_mapper}
\input{6_discussion}
\input{7_related_work}
\input{8_conclusion}

\bibliographystyle{ACM-Reference-Format}
\bibliography{paper}

\end{document}

%% file: urls.tex
\newcommand{\URLClassifier}{\url{https://github.com/x4dx65/SpeedReader.git}\xspace}
\newcommand{\URLLabels}{\url{https://github.com/x4dx65/SpeedReader/blob/master/labels/labels.csv}\xspace}
\newcommand{\URLLabelsGuide}{\url{https://github.com/x4dx65/SpeedReader/blob/master/labels/labels-legend.txt}\xspace}

%% file: abbr.tex
\newcommand{\OSN}{online social network\xspace}
\newcommand{\OSNs}{online social network\xspace}
\newcommand{\ToolName}{\texttt{SpeedReader}\xspace}
\newcommand{\MoeClassifier}{SpeedReader Classifier\xspace}

\newcommand{\RM}{reader mode\xspace}
\newcommand{\uRM}{Reader mode\xspace}
\newcommand{\uuRM}{Reader Mode\xspace}

\newcommand{\RMs}{reader modes\xspace}

\newcommand{\StepOne}{classification\xspace}
\newcommand{\uStepOne}{Classification\xspace}
\newcommand{\StepTwo}{tree transduction\xspace}
\newcommand{\uStepTwo}{Tree transduction\xspace}
\newcommand{\uuStepTwo}{Tree Transduction\xspace}

\newcommand{\RABy}{readability\xspace}
\newcommand{\RMA}{readable\xspace}
\newcommand{\uRMA}{Readable\xspace}

\newcommand{\SRV}{Safari Reader View\xspace}
\newcommand{\DD}{DOM Distiller\xspace}
\newcommand{\RJS}{Readability.js\xspace}
\newcommand{\BP}{BoilerPipe\xspace}

\newcommand{\JS}{JavaScript\xspace}

\newcommand{\EL}{EasyList\xspace}
\newcommand{\EP}{EasyPrivacy\xspace}

%% file: data.tex
\newcommand{\ClassifierReadabilityJSPrecision}{68\%\xspace}
\newcommand{\ClassifierReadabilityJSRecall}{85\%\xspace}
\newcommand{\ClassifierDomDistillerPrecision}{90\%\xspace}
\newcommand{\ClassifierDomDistillerRecall}{75\%\xspace}
\newcommand{\ClassifierMoePrecision}{91\%\xspace}
\newcommand{\ClassifierMoeRecall}{87\%\xspace}
\newcommand{\ClassifierMoeAccuracy}{91\%\xspace}

\newcommand{\ClassifierTrainingLandingNum}{932\xspace}
\newcommand{\ClassifierTrainingLandingPct}{1.5\%\xspace}
\newcommand{\ClassifierTrainingRandNum}{956\xspace}
\newcommand{\ClassifierTrainingRandPct}{21.8\%\xspace}
\newcommand{\ClassifierTrainingArticleNum}{945\xspace}
\newcommand{\ClassifierTrainingArticlePct}{92.9\%\xspace}
\newcommand{\ClassifierTrainingSubSetSize}{1,000\xspace}
\newcommand{\ClassifierTrainingNonRespondingSitesNum}{167\xspace}
\newcommand{\ClassifierTrainingInitialNum}{3,000\xspace}
\newcommand{\ClassifierTrainingTotalNum}{2,833\xspace}
\newcommand{\ClassifierTrainingTotalPct}{38.8\%\xspace}

\newcommand{\MapperEvalDatasetDate}{17-20 October 2018\xspace}
\newcommand{\MapperEvalFetchedPages}{91,439\xspace}
\newcommand{\MapperEvalFetchedDomains}{10,000\xspace}
\newcommand{\MapperEvalDatasetNumPages}{19,765\xspace}
\newcommand{\MapperEvalReadableNum}{\MapperEvalDatasetNumPages\xspace}
\newcommand{\MapperEvalReadablePct}{21.62\%\xspace}
\newcommand{\MapperEvalDatasetNumDomains}{4,931\xspace}

\newcommand{\ApplicabilityNumUniquePages}{83,894\xspace}
\newcommand{\ApplicabilityNumUniquePagesReadable}{18,457\xspace}
\newcommand{\ApplicabilityNumUniquePagesReadablePct}{22.0\%\xspace}

\newcommand{\ApplicabilityOSNPagesReddit}{3,035\xspace}
\newcommand{\ApplicabilityOSNPagesRedditReadable}{1,260\xspace}
\newcommand{\ApplicabilityOSNPagesRedditReadablePct}{41.51\%\xspace}
\newcommand{\ApplicabilityOSNPagesTwitter}{494\xspace}
\newcommand{\ApplicabilityOSNPagesTwitterReadable}{276\xspace}
\newcommand{\ApplicabilityOSNPagesTwitterReadablePct}{31.2\%\xspace}
\newcommand{\ApplicabilityOSNPagesRSS}{506\xspace}
\newcommand{\ApplicabilityOSNPagesRSSReadable}{331\xspace}
\newcommand{\ApplicabilityOSNPagesRSSReadablePct}{65\%\xspace}
\newcommand{\ApplicabilityOSNPagesTotal}{4,035\xspace}
\newcommand{\ApplicabilityOSNPagesTotalReadable}{1,867\xspace}
\newcommand{\ApplicabilityOSNPagesTotalReadablePct}{46.27\%\xspace}

\newcommand{\ApplicabilityAlexaFiveK}{42,986\xspace}
\newcommand{\ApplicabilityAlexaFiveKReadable}{9,653\xspace}
\newcommand{\ApplicabilityAlexaFiveKReadablePct}{22.5\%\xspace}
\newcommand{\ApplicabilityAlexaUnpopular}{40,908\xspace}
\newcommand{\ApplicabilityAlexaUnpopularReadable}{8,794\xspace}
\newcommand{\ApplicabilityAlexaUnpopularReadablePct}{21.5\%\xspace}

\newcommand{\ClassifierMeanTime}{2.8\xspace}
\newcommand{\ClassifierMedianTime}{1.9\xspace}
\newcommand{\InitHtmlTraceMeanTime}{22.3\xspace}
\newcommand{\InitHtmlTraceMedianTime}{15.5\xspace}
\newcommand{\InitHtmlCurlMeanTime}{1,372\xspace}
\newcommand{\InitHtmlCurlMedianTime}{652\xspace}
\newcommand{\InitHtmlCurlMobileMeanTime}{4,023\xspace}
\newcommand{\InitHtmlCurlMobileMedianTime}{2,516\xspace}

\newcommand{\MapperDefaultMeanRequests}{144\xspace}
\newcommand{\MapperDefaultMeanKBFetched}{2,283\xspace}
\newcommand{\MapperDefaultMeanMBMem}{197\xspace}
\newcommand{\MapperDefaultMeanRenderMs}{1,813\xspace}
\newcommand{\MapperDefaultMeanTP}{117\xspace}
\newcommand{\MapperDefaultMeanScripts}{83\xspace}
\newcommand{\MapperDefaultMeanLabeled}{63\xspace}
\newcommand{\MapperDefaultMedianRequests}{91\xspace}
\newcommand{\MapperDefaultMedianKBFetched}{1,461\xspace}
\newcommand{\MapperDefaultMedianMBMem}{174\xspace}
\newcommand{\MapperDefaultMedianRenderMs}{1,069\xspace}
\newcommand{\MapperDefaultMedianTP}{63\xspace}
\newcommand{\MapperDefaultMedianScripts}{51\xspace}
\newcommand{\MapperDefaultMedianLabeled}{24\xspace}

\newcommand{\MapperDomDistillerMeanRequests}{5\xspace}
\newcommand{\MapperDomDistillerMeanKBFetched}{186\xspace}
\newcommand{\MapperDomDistillerMeanMBMem}{84\xspace}
\newcommand{\MapperDomDistillerMeanRenderMs}{550\xspace}
\newcommand{\MapperDomDistillerMeanTP}{3\xspace}
\newcommand{\MapperDomDistillerMeanScripts}{0\xspace}
\newcommand{\MapperDomDistillerMeanLabeled}{0\xspace}
\newcommand{\MapperDomDistillerMedianRequests}{2\xspace}
\newcommand{\MapperDomDistillerMedianKBFetched}{61\xspace}
\newcommand{\MapperDomDistillerMedianMBMem}{79\xspace}
\newcommand{\MapperDomDistillerMedianRenderMs}{63\xspace}
\newcommand{\MapperDomDistillerMedianTP}{1\xspace}
\newcommand{\MapperDomDistillerMedianScripts}{0\xspace}
\newcommand{\MapperDomDistillerMedianLabeled}{0\xspace}

\newcommand{\MapperReadabilityJSMeanRequests}{5\xspace}
\newcommand{\MapperReadabilityJSMeanKBFetched}{186\xspace}
\newcommand{\MapperReadabilityJSMeanMBMem}{85\xspace}
\newcommand{\MapperReadabilityJSMeanRenderMs}{583\xspace}
\newcommand{\MapperReadabilityJSMeanTP}{3\xspace}
\newcommand{\MapperReadabilityJSMeanScripts}{0\xspace}
\newcommand{\MapperReadabilityJSMeanLabeled}{0\xspace}
\newcommand{\MapperReadabilityJSMedianRequests}{2\xspace}
\newcommand{\MapperReadabilityJSMedianKBFetched}{61\xspace}
\newcommand{\MapperReadabilityJSMedianMBMem}{79\xspace}
\newcommand{\MapperReadabilityJSMedianRenderMs}{68\xspace}
\newcommand{\MapperReadabilityJSMedianTP}{1\xspace}
\newcommand{\MapperReadabilityJSMedianScripts}{0\xspace}
\newcommand{\MapperReadabilityJSMedianLabeled}{0\xspace}

\newcommand{\MapperBoilerPlateMeanRequests}{2\xspace}
\newcommand{\MapperBoilerPlateMeanKBFetched}{101\xspace}
\newcommand{\MapperBoilerPlateMeanMBMem}{81\xspace}
\newcommand{\MapperBoilerPlateMeanRenderMs}{545\xspace}
\newcommand{\MapperBoilerPlateMeanTP}{1\xspace}
\newcommand{\MapperBoilerPlateMeanScripts}{0\xspace}
\newcommand{\MapperBoilerPlateMeanLabeled}{0\xspace}
\newcommand{\MapperBoilerPlateMedianRequests}{2\xspace}
\newcommand{\MapperBoilerPlateMedianKBFetched}{61\xspace}
\newcommand{\MapperBoilerPlateMedianMBMem}{77\xspace}
\newcommand{\MapperBoilerPlateMedianRenderMs}{44\xspace}
\newcommand{\MapperBoilerPlateMedianTP}{1\xspace}
\newcommand{\MapperBoilerPlateMedianScripts}{0\xspace}
\newcommand{\MapperBoilerPlateMedianLabeled}{0\xspace}

\newcommand{\MapperReadabilityJSMeanTPSaved}{115\xspace}

\newcommand{\MapperReadabilityJSMeanLabeledSaved}{64\xspace}

\newcommand{\MapperReadabilityJSMeanRequestsGain}{51\xspace}
\newcommand{\MapperReadabilityJSMedianRequestsGain}{28\xspace}
\newcommand{\MapperReadabilityJSMeanKBFetchedGain}{84\xspace}
\newcommand{\MapperReadabilityJSMedianKBFetchedGain}{24\xspace}
\newcommand{\MapperReadabilityJSMeanMBMemGain}{2.4\xspace}
\newcommand{\MapperReadabilityJSMedianMBMemGain}{2.1\xspace}
\newcommand{\MapperReadabilityJSMeanRenderMsGain}{20\xspace}
\newcommand{\MapperReadabilityJSMedianRenderMsGain}{11\xspace}
\newcommand{\MapperDomDistillerMeanRequestsGain}{52\xspace}
\newcommand{\MapperDomDistillerMedianRequestsGain}{32\xspace}
\newcommand{\MapperDomDistillerMeanKBFetchedGain}{84\xspace}
\newcommand{\MapperDomDistillerMedianKBFetchedGain}{24\xspace}
\newcommand{\MapperDomDistillerMeanMBMemGain}{2.4\xspace}
\newcommand{\MapperDomDistillerMedianMBMemGain}{2.1\xspace}
\newcommand{\MapperDomDistillerMeanRenderMsGain}{21\xspace}
\newcommand{\MapperDomDistillerMedianRenderMsGain}{12\xspace}
\newcommand{\MapperBoilerPlateMeanRequestsGain}{77\xspace}
\newcommand{\MapperBoilerPlateMedianRequestsGain}{48\xspace}
\newcommand{\MapperBoilerPlateMeanKBFetchedGain}{84\xspace}
\newcommand{\MapperBoilerPlateMedianKBFetchedGain}{24\xspace}
\newcommand{\MapperBoilerPlateMeanMBMemGain}{2.4\xspace}
\newcommand{\MapperBoilerPlateMedianMBMemGain}{2.1\xspace}
\newcommand{\MapperBoilerPlateMeanRenderMsGain}{27\xspace}
\newcommand{\MapperBoilerPlateMedianRenderMsGain}{15\xspace}

\newcommand{\MapperMeanRenderMsGainMin}{20}
\newcommand{\MapperMeanRenderMsGainMax}{27}

\newcommand{\MapperMeanKBFetchedGainMax}{84}

\newcommand{\MapperMeanMBMemGainMax}{2.4}
\newcommand{\MapperMeanRequestsGainMin}{51}
\newcommand{\MapperMeanRequestsGainMax}{77}

%% file: commands.tex
\ifDRAFT
  \newcommand{\fix}[1]{\textcolor{red}{#1}}
  \newcommand{\todo}[1]{{\color{red}\bf\em TODO:\/\@#1}}
  
\else
  \newcommand{\fix}[1]{{}}
  \newcommand{\todo}[1]{{}}
  
\fi

\newcommand{\point}[1]{\par\smallskip\noindent\textbf{#1.}}
\newcommand{\etal}{\textit{et al.}}

\newcolumntype{+}{>{\global\let\currentrowstyle\relax}}
\newcolumntype{^}{>{\currentrowstyle}}
\newcommand{\rowstyle}[1]{\gdef\currentrowstyle{#1}%
#1\ignorespaces
}

%% file: 0_abstract.tex
\begin{abstract}

Most popular web browsers include ``reader modes'' that improve the user
    experience     by removing un-useful page elements. Reader modes reformat the
    page to hide     elements that are not related to the page's main content.
    Such page     elements include site navigation, advertising related videos and
    images,     and most JavaScript. The intended end result is that users can
    enjoy the     content they are interested in, without distraction.

In this work, we consider whether the ``reader mode'' can be widened     to
    also provide performance and privacy improvements. Instead of its use as a
    post-render feature to clean up the clutter on a page we propose \ToolName as
    an alternative multistep pipeline that is part of the rendering pipeline. Once
    the tool decides during the initial phase of a page load that a page is
    suitable for reader mode use, it directly applies document tree
    translation \textit{before the page is rendered}.

Based on our measurements, we believe that \ToolName can be continuously
    enabled in order to drastically improve end-user experience, especially on
    slower mobile connections. Combined with our approach to     predicting which
    pages should be rendered in reader mode with \ClassifierMoeAccuracy accuracy,
    it achieves drastic speedups and bandwidth reductions of up to
    \MapperMeanRenderMsGainMax$\times$ and \MapperMeanKBFetchedGainMax$\times$
    respectively on average.     We further find that our novel ``reader
    mode'' approach brings with it significant privacy improvements to users. Our
    approach effectively removes all commonly recognized trackers, issuing
    \MapperReadabilityJSMeanTPSaved fewer requests to     third parties, and
    interacts with \MapperReadabilityJSMeanLabeledSaved     fewer trackers on
    average, on transformed pages.

% We show that our approach is able to match expert labels of whether there is a
%     ``readable'' subset of the page with \ClassifierMoeAccuracy accuracy.  We
%     find that our approach has significant performance implications, and is
%     able to approximate existing ``reader mode'' levels of usability while
%     fetching on average \MapperReadabilityJSMeanKBFetchedSaved KB less data, using
%     \MapperReadabilityJSMeanMBMemSaved MB less memory, and saving
%     \MapperReadabilityJSMeanRenderMsSaved ms to render on applicable pages.  We
%     further find that our novel ``reader mode'' approach brings with it
%     significant privacy improvements to users when compared to both default
%     browsing, and existing reader mode tools.  Our approach issues
%     \MapperReadabilityJSMeanTPSaved fewer requests to third parties, and
%     interacts with \MapperReadabilityJSMeanLabeledSaved fewer ad and tracking
%     related resources, on applicable pages.  We also implement our proposed
%     ``reader mode'' strategy as a patch to Chromium.
\end{abstract}

%% file: 1_introduction.tex
\section{Introduction}
\label{sec:introduction}
``Web bloat'' is a colloquial term that describes the trend in websites to
accumulate size and visual complexity over time. The phenomena has been measured
in many dimensions, including total page
size~\cite{butkiewicz2011understanding}, page load
time~\cite{Bouch:2000:QEB:332040.332447, xiao:2013, xiao:2016}, memory
needed~\cite{merzdovnik2017block}, number of network requests~\cite{libert:2015,
goel2017measuring}, amount of scripts executed
~\cite{Ratanaworabhan:2010:JCB:1863166.1863169, Snyder:2016:BFU:2987443.2987466,
Kumar:2017:SCI:3038912.3052686, Nikiforakis:2012:YYI:2382196.2382274} and third
parties contacted~\cite{Krishnamurthy:2009:PDW:1526709.1526782,
Kumar:2017:SCI:3038912.3052686, libert:2015}. This work suggests that growth in
page size and complexity is outpacing improvements in device hardware. All of
this has a predictably negative impact on user experience.

Web users and browser vendors have reacted to this ``bloat'' in a variety of
ways, all partially helpful, but with significant downsides.

Ad and tracking blockers are a popular and useful tool for reducing the size
complexity of sites.  Prior work has shown that these tools can be effective in
reducing privacy leaks~\cite{Merzdovnik2017}, network use, and extend device
memory life.  Such tools are inherently limited in the scope of improvements
they can achieve. While these lists are large~\cite{vastel2018easylist}, they
are small as a proportion of all URLs on the web.  Similarly, while these lists
are updated often, they are updated slowly compared to URL updates on the
web.

Similarly, ``reader mode'' tools, provided in many popular browsers and
browser extensions, are an effort to reduce the growing visual complexity
of web sites. Such tools attempt to extract the subset of page content
useful to users, and remove advertising, animations, boiler plate code, and
other non-core content.  Current ``reader modes'' do not provide the user with 
resource savings since the referenced resources have already been fetched and
rendered.  The growth and popularity of such tools suggest they are useful
to browser users, looking to address the problem of page clutter and visual
``bloat''.

In this work, we propose a novel strategy called \textbf{\ToolName} for dealing
with resource and bloat on websites. Our technique provides a user experience
similar to existing ``reader mode'' tools, but with network, performance, and
privacy improvements that exceed existing ad and tracking blocking tools, on a
significant portion of websites.  Significantly, \ToolName differs from
existing deployed \RM tools by operating \textit{before page rendering}, which
allows it to determine which resources are needed for the page's core content
before fetching.  

\point{How we achieve speedups} \ToolName achieves its performance improvements
through a two-step pipeline:

\begin{enumerate}
\item \ToolName uses a classifier to determine whether
    there is a \RMA subset of the initial, fetched page HTML.  This classifier is
    trained on a labeled corpus of \ClassifierTrainingTotalNum websites, and is
    described in detail in Section~\ref{sec:classifier}, and determines whether
    a page can be display in \RM with \ClassifierMoeAccuracy accuracy.
\item If the classifier has determined that the page is \RMA, \ToolName extracts
    the \RMA subset of document \textit{before rendering}, using a variety of
    heuristics developed in prior research~\cite{kohlschutter2010boilerplate}
    and browser vendors~\cite{domdistiller, readabiltiyjs}, and passes the
    simplified, \RM document to the browser's render layer.  This tree
    translation step is described in Section~\ref{sec:mapper}.
\end{enumerate}

\point{Deployment} Combined with a highly accurate classifier of ``readable''
pages, the drastic improvements in performance, reduction in bandwidth use and
elimination of trackers in reader mode make the approach practical for
continuous use. We therefore propose \ToolName as a sticky feature that a user
can toggle to be always on. This approximates the experience of using an e-book
reader, but with strengths of content availability on the web. It is also a
suitable strategy for content prerendering or prefetching that could be
implemented by web browser vendors, automatically delivering graceful
performance degradation in poor connectivity areas or on underpowered mobile
devices until the rest of the page content can be fetched for a complete render.

\point{Contributions}
\begin{itemize}
\item \textbf{Novel approach to \uuRM} - combining machine-learning
    driven approach to checking whether content can be transformed to
    text-focused representation for end-user consumption.
\item \textbf{Applicability} - we demonstrate that a
    \ApplicabilityNumUniquePagesReadablePct of web pages are convertible to
    \RM style in a dataset of pages reported popular by Alexa. We
    further demonstrate that \ApplicabilityOSNPagesTotalReadablePct of pages
    shared on social networks are \textit{readable}.
\item \textbf{Privacy} - we demonstrate that using \RM in the
    proposed design provides superior privacy protection, effectively removing
    all trackers from the tested pages, and dramatically reducing communication
    with third-parties.
\item \textbf{Ad Blocking} - we show that our unique \RM approach
    blocks ads \textit{at least as well} as existing ad blocking
    tools, blocking 100\% of resources labeled as advertising
    related by \EL in a crawl of \MapperEvalFetchedPages pages, without the
    needed to use hand curated, hard-coded filter lists.
\item \textbf{Speed} - we find that the lightweight nature of \RM
    content results in huge performance gains, with up to
    \MapperMeanRenderMsGainMax$\times$ page load time speedup on average,
    together with up to \MapperMeanKBFetchedGainMax$\times$ bandwidth and
    \MapperMeanMBMemGainMax$\times$ memory reduction on average.
\end{itemize}

\point{Paper organization} The rest of this paper is structured as follows.
Section~\ref{sec:background} provides background information to place \ToolName
in context. Section~\ref{sec:classifier} describes the design, evaluation and
accuracy of the classifying step in the \ToolName pipeline, and Section
\ref{sec:mapper} gives the design and evaluation of the \RM extraction step in
the \ToolName pipeline. Section~\ref{sec:applicability} measures how many
websites user encounter is amenable to \ToolName, under several browser use
scenarios. Section~\ref{sec:discussion} provides some discussion for how our
findings can inform future readability, privacy and performance work, Section
\ref{sec:related-work} places this work in the context of prior research, and
Section~\ref{sec:conclusion} concludes.

%% file: 2_background.tex
\section{Background}
\label{sec:background}

\subsection{Terminology}
This subsection presents several terms that are not standardized.  We present
them up front, to ease the understanding of the rest of the work.

\point{\uRM}
We use the term ``\RM'' to describe any tool that attempts to extract
a useful subset of a website for a simplified presentation.  These tools
can be either included in the web browser by the browser vendor, added by
users through browser extensions, or provided by third parties as a web service.
Our use of the term ``\RM'' is generic to the concept, and should not be
confused with any specific tool.

\point{\uStepOne and transduction}
\uRM tools generally include both a technique for determining whether a page is
\RMA, which we refer to as ``\StepOne'', and a strategy for
converting the initial HTML tree into a simplified \RM tree, which we refer to
as ``\StepTwo''.  Though most \RM tools include both steps within a single
tool or library, they are conceptually distinct.

\point{\uRMA}
We use the term ``\RMA'' to describe whether a web page contains a
subset of content that would be useful to display in a \RM presentation.  \RM
presentation works best on pages that are text and image focused, and that are
mostly static (i.e. few interactive page elements).  Examples of such \RMA pages
include articles, blog posts, and news sites. \uRM presentation does not
work well on web sites that are highly interactive, or when the a page's
structure is significant to the page's content.  Examples of such non-\RMA
pages include web applications (e.g. Google Mail, Google Maps) or pages that
are indexes of other content.

\subsection{Existing Reader Modes}
\label{sec:background:existing}
Several popular web browsers include \RMs designed to simplify a page's
presentation, so that browser users can read the page's contents, without
visual clutter.  Most \RM tools have the same general goal, to display only
the text, images and resources a user needs to achieve their goals on the
web site, without the distraction of advertisements, page animations, and 
unnecessary page boiler plate (e.g. footers, page navigation, comments).

In this section, we give a brief description of several existing \RM tools,
how they're deployed by their authors, and how they are used in the evaluations
given in the rest of this paper.

\point{\RJS}
\RJS~\cite{readabiltiyjs} is an open source \RM library, implemented in \JS.
It is maintained by Mozilla, and is used for the \RM function in Firefox.  The
code is closely related to ``Readability''~\cite{arc90readability}, a open
sourced library developed by Arc90 and used for their now-defunct
\url{readability.com} web service.  \uStepOne works by looking for the
element on the page with the highest density of text and link nodes.  If the
number and density of text and link nodes in that element exceed a given
threshold, the library treats the page as \RMA. \uStepTwo works by 
normalizing the contents of the text-and-link dense element (to remove styling
and other mark up), looking for near-by images for inclusion, and using text
patterns in the document that identify the page's author, source and
publication date.

Significant to \ToolName, \RJS does not consider any display or presentation
information when performing either the \StepOne or \StepTwo steps.  This means
that the page does not need to be loaded and rendered to generate a \RM
presentation (though in practice Firefox does not use this library in this way).

\point{\SRV}
\SRV is a \JS library that implements the \RM presentation in
Safari.  Like \RJS, it is also a fork from Arc90's ``Readability'', though
Apple has changed how the library works in significant ways.  In addition to
looking for elements with high text and anchor density, \SRV
also uses presentation-level heuristics, including where elements appear on
the page and what elements are hidden from display by default.  Relevant to
\ToolName, this means that \SRV must load a page and at least some of its
resources (e.g. images, CSS, \JS) to perform either the \StepOne or \StepTwo
level decisions.  Because significant portions of \SRV require a document
be fetched and rendered before being evaluated, we do not consider it further
in this work (for reasons that are detailed in Sections~\ref{sec:classifier} and
\ref{sec:mapper}).

\point{\BP}
\BP is an academic research project from Kohlsch{\"u}tter 
\etal~\cite{kohlschutter2010boilerplate}, and is implemented in Java. \BP has not
been deployed directly by any browser vendors.  \BP does not provide
functionality for (readability) \StepOne, and assumes that any HTML document contains a
\RMA subset.  For \StepTwo, \BP considers number of words and link density
features. Like \RJS, it does not require a browser to load and render a page
in order to do \RM extraction. Their analysis reveals a strong 
correlation between short text and boilerplate, as well as long text and actual content text
(of the textual content) on the Web. Using features with low calculation cost 
such as number of words enables \BP to lower the overhead while maintaining high 
accuracy.

\point{\DD}
\DD is a \JS and C++ library maintained by Google, and used to implement \RM
in recent versions of Chrome.  The project is based on \BP, though has been
significantly changed by Google.  The \StepOne step in \DD uses a classifier
based approach, and considers features such as whether the page's URL contains
certain keywords (e.g. ``forum'', ``.php'', ``index''), if the page's markup
contains Facebook open graph, Google AMP, identifiers, or the number of ``/''
characters used in the URL's path, in addition to the text-and-link density
measures used by \RJS.  At a high level, the \StepTwo step also looks
text-and-link dense element in the page, as well as special-cased embedded
elements, such as YouTube or Vimeo videos.

\DD considers some render-level information in both the \StepOne and \StepTwo
steps.  For example, any elements that are hidden from display are not included
in the text-and-link density measurements.  These render-level checks are
a small part of \DD's strategy.  We modified \DD to remove these display level
checks, so that \DD could be applied to prerendered HTML documents.
We note that the evaluation of \DD in this work uses this
modified version of \DD, and not the version that Google ships with Chrome.
We expect this modification to have minimal effects on the discussed
measurements, but draw the reader's attention to this change for
completeness.

\subsection{Comparison to \ToolName}

The \RM functionality shipped with all current major browsers is applied after
the document is fully fetched and rendered.\footnote{While \RJS
does not require that the page be rendered before making \RM evaluations, in
practice Firefox does not expose \RM functionality to users until after the
page is fetched and loaded.} This greatly restricts the the possible
performance, network and privacy improvements existing \RMs can
achieve.  In fact, in some \RM implementations we measured, using \RMs
\textit{increased} the amount of network used, as some resources were fetched
twice, i.e. once for the initial page loading, and then again when presenting
images in the \RM display.

\begin{figure}[t]
  \centering
  \includegraphics[width=0.75\columnwidth]{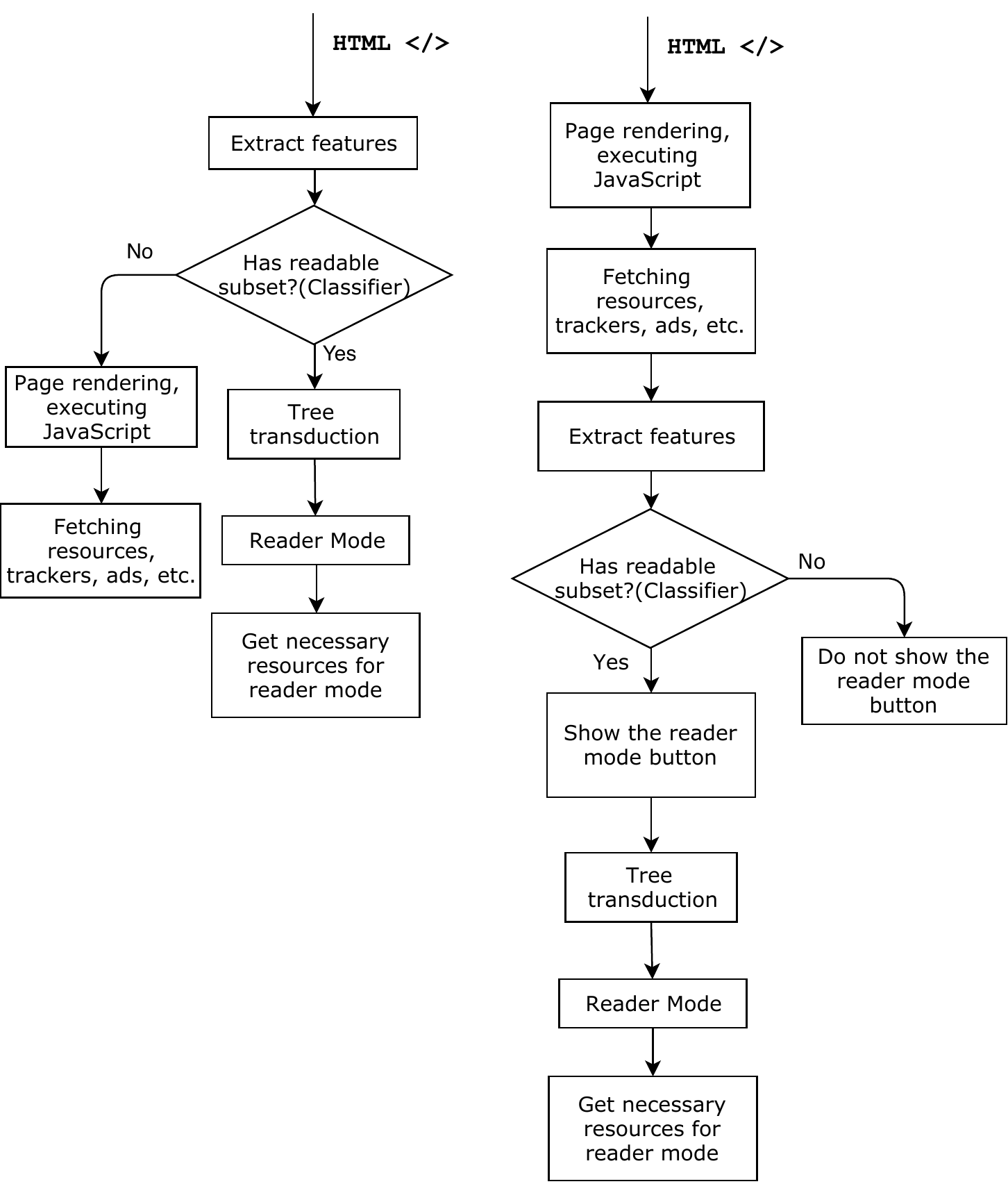}
  \caption{Comparison of SpeedReader (left) with other existing reader modes (right)}
  \label{fig:background:existing:comparison}
\end{figure}

Most significantly, \ToolName differs from existing \RM techniques in that it
is implemented strictly before the display, rendering, and resource-fetching steps
 in the browser. \ToolName can therefore be thought of as a function that sits between the
browser's network layer (i.e. takes as input the initial HTML document), and
returns either the received HTML (when there is no \RMA subset), or a greatly
simplified HTML document, representing the \RM presentation (when there is a
\RMA subset). Figure ~\ref{fig:background:existing:comparison} provides a high
level comparison of how \ToolName functions, compared to existing \RMs.

\begin{figure}[t]
  \centering
  \includegraphics[width=.9\columnwidth]{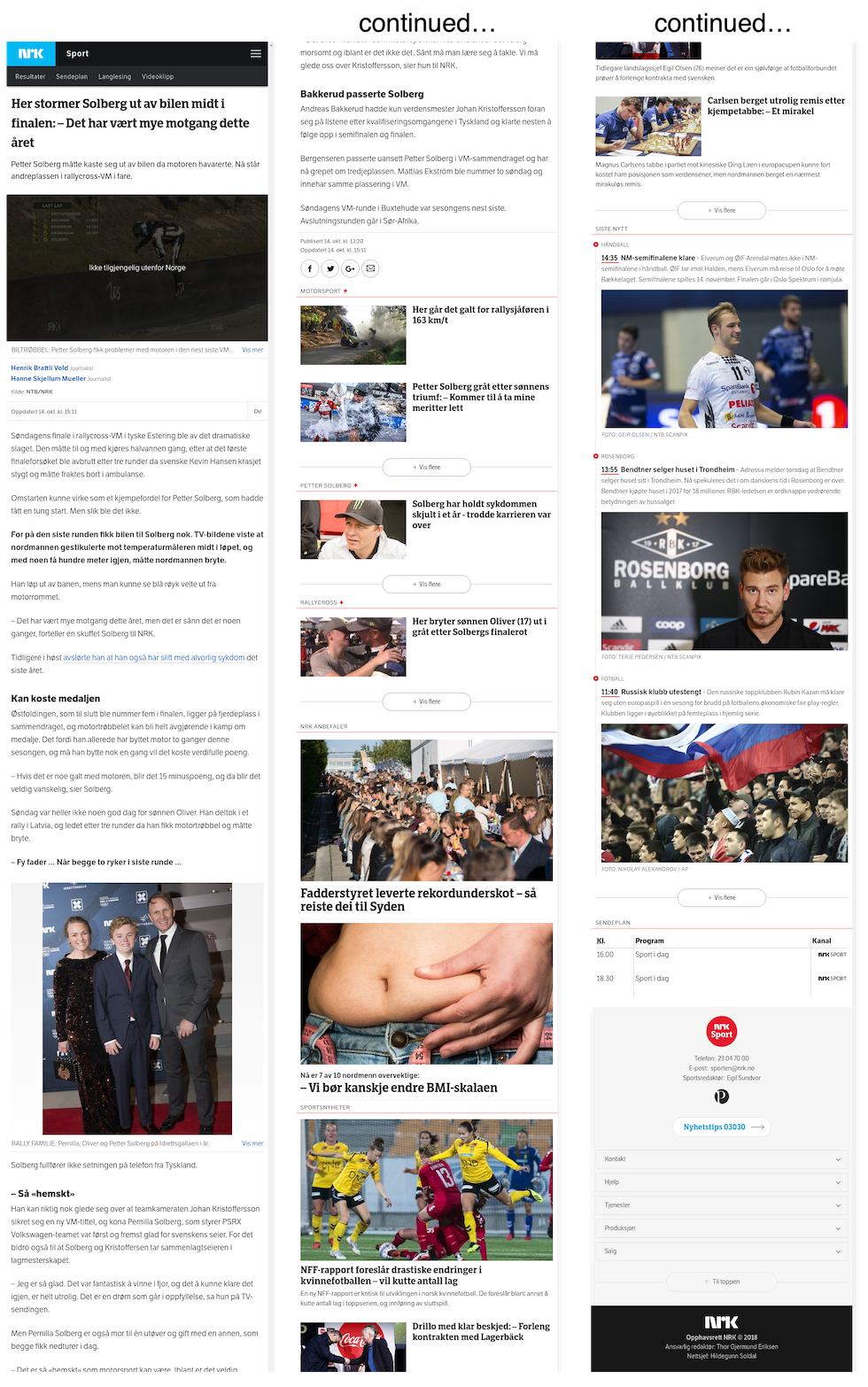}
  \caption{An example page loaded with Google Chrome browser with no modifications}
  \label{fig:background:defaultRendering}
\end{figure}

\begin{figure}[t]
  \centering
  \includegraphics[width=.8\columnwidth]{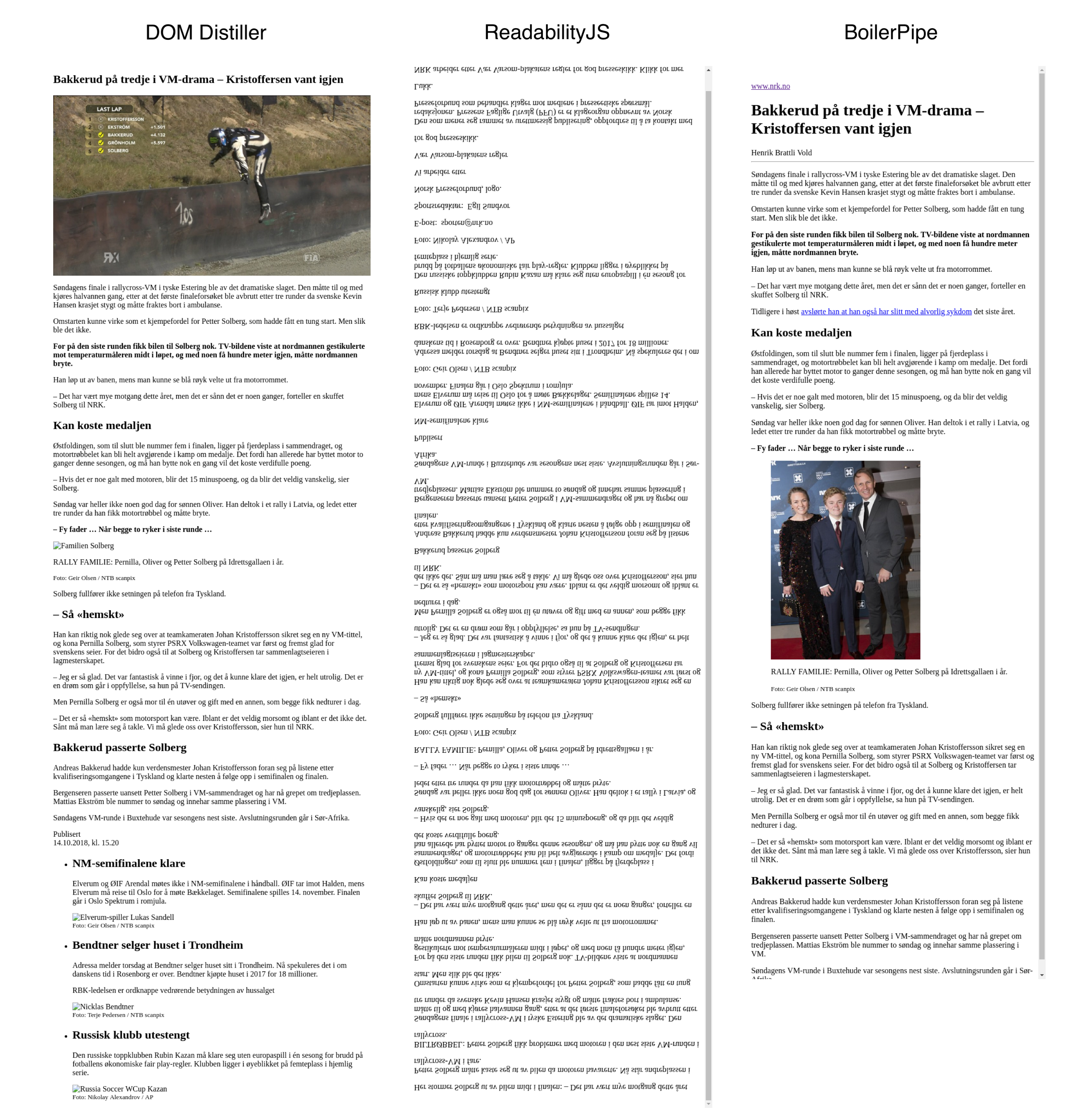}
  \caption{The example page transformed with each of the evaluated \ToolName transducers}
  \label{fig:background:renderingComparison}
\end{figure}

That \ToolName only considers features available in the initial HTML and URL
enables \ToolName to achieve performance orders of magnitude above existing
approaches. Figure~\ref{fig:background:defaultRendering} provides a strawman
example of a news page as delivered to a standard client: including portal
branding and content, but also a range of links to different articles, images
and trackers, for a total of 2.7MB of data transferred and 53 scripts
executed. Figure~\ref{fig:background:renderingComparison}
demonstrates the functionality of \ToolName when applying existing \RM
transducers to just the initial HTML document. Therefore, for documents
\ToolName determines are \RMA, the sources of \ToolName improvements include:

\begin{itemize}
  \item Never fetching or executing script or CSS.
  \item Fetching far fewer images or videos (since images and videos not 
        core to the page's presentation are never retrieved).
  \item Performing network requests to far fewer third parties (zero, in
        the common case).
  \item Saving processing power from not rendering animations, videos
        or complex layout operations, since \RM presentations
        of page content are generally simple. 
\end{itemize}
The above are just some of the ways that \ToolName is able to achieve dramatic
performance improvements.  The following sections describe how \ToolName's
\StepOne and \StepTwo steps were designed and evaluated, and what percentage of
websites are amenable to \ToolName's approach.

%% file: 3_classifier.tex
\section{Page \uStepOne}
\label{sec:classifier}
\ToolName uses a two stage pipeline for generating \RM versions of websites.
This section presents the design and evaluation of the first half of the
pipeline, the \StepOne step.

% This section proceeds by first describing the design of the \RABy classifier in
% \ToolName, then presents a measurement of how the classifier performs on a 
% hand-labeled data set, both on its own and compared against other deployed
% \RM classification strategies.  Finally, we present a measurement of the
% rendering delay imposed by our classification technique and find it to be
% acceptable.

\subsection{Classifier Design}
\label{sec:classifier:speedreader}
\input{classifier-features}

The \StepOne step of \ToolName uses a random forest classifier, trained on a
hand-labeled data set of \ClassifierTrainingTotalNum websites.  Our classifier
takes as input a string, depicting a HTML document, and returns a boolean label
of whether there is a \RMA subset of the document.  We note that the
input to the classifier is the initial HTML returned by the server, and not the
final state of the website after \JS execution.

Our classifier is designed to execute quickly, since document rendering is
delayed during classification. The classifier considers 21 features, each
selected to be extractable quickly. Selected features include the number
of text nodes, number of words, the presence of Facebook open graph or Google
AMP markup, and counts for a variety of other tags. 

Table~\ref{table:classifier-features} presents the full set of features considered
by the classifier.  The source code for the classifier is publicly available as well
\footnote{\URLClassifier}.

\subsection{Classifier Accuracy}
\input{classifier-accuracy-dataset.tex}
The goal of the classifier in \ToolName is to predict whether the end result
of a page's fetching and execution will result in a \RMA page, based on the
initial HTML of the page.  This section describes the data set we used to both
train the \ToolName classifier, and to evaluate its accuracy against existing
popular, deployed \RM tools.

\point{Data Set}
To assess the accuracy of our classifier, we first gathered
\ClassifierTrainingTotalNum websites, summarized in Table
\ref{table:classifier-accuracy-dataset}.  Our data set is made up of three
smaller sets of crawl data, each containing \ClassifierTrainingSubSetSize URLs,
each meant to focus on a different kind of page, with a different expected
distribution of \RABy.  \ClassifierTrainingSubSetSize pages were
URLs selected from the RSS feeds of popular news sites (e.g. 
The New York Times, ArsTechnica), which we expected to be freqently \RMA.
The second \ClassifierTrainingSubSetSize pages were the landing pages from the
Alexa 1K, which we expected to rarely be readable.  The final
\ClassifierTrainingSubSetSize pages were selected randomly from non-landing
pages linked from the landing pages of the Alexa 5K, which we expected to be
occasionally \RMA.

We built a crawler that, given a URL, recorded both the initial HTML
response, and a screenshot of the final rendered page (i.e. after all resources
had been fetched and rendered, and after \JS had executed).  We applied our
crawler to each of the \ClassifierTrainingInitialNum selected URLs.
\ClassifierTrainingNonRespondingSitesNum pages did not respond to our crawler,
accounting for the difference between the \ClassifierTrainingInitialNum selected
URLs and the \ClassifierTrainingTotalNum pages in our data set.

Finally, we manually considered each of the final page screenshots, and gave each
a boolean label of whether there was a subset of page content that was \RMA.
We considered a page \RMA if it met the following criteria:

\begin{enumerate}
  \item The primary utility of the page was its text and image content (i.e. not
        interactive functionality).
  \item The page contained a subset of content that was useful, without
        being sensitive to its placement on the page.
  \item The usefulness of the page's content was not dependent on its specific
        presentation or layout on the website.
\end{enumerate}
This meant that single page applications, index pages, and pages with complex
layout were generally labeled as not-\RMA, while pages with generally static
content, and lots of text and content-depicting media, were generally
labeled \RMA. We also share our labeled data\footnote{\URLLabels}, and a
guide to the meaning behind the labels\footnote{\URLLabelsGuide}, to make
our results transparent and reproducible.

\input{classifier-accuracy.tex}

\point{Evaluation}
We evaluated our classifier on our hand labeled corpus of
\ClassifierTrainingTotalNum websites, performing a standard ten-fold cross-validation.
For comparison sake, we also evaluated the accuracy of the
classification functionality in \RJS and our modified version of \DD when
applied to the same data set, to judge their ability to predict the final \RABy
state of each document, given its initial HTML.

We note that \RJS is designed to be used this way, but that this prediction
point is slightly different that how \DD is deployed in Chrome. In Chrome, \DD
labels a page as \RMA based on its final rendered state. This evaluation of
\DD's \StepOne capabilities should therefore not be seen as an evaluation of
\DD's overall quality, but only its ability to achieve the kinds of
optimizations sought by \ToolName.

We find that \ToolName is able to classify the final \RMA state of HTML
documents better than existing tools.
Table~\ref{table:classifier-accuracy} presents the results of this measurement.
As the table shows, \ToolName strictly outperforms the \StepOne capabilities of
both \DD and \RJS.  \DD has a higher false positive rate for the task, while
\RJS has a higher false negative rate.

\subsection{Classifier Usability}
\label{sec:applicability}

\point{Problem Statement}
Our classifier operates on complete HTML documents, before they are rendered.
As a result, the browser is not able to render the document until the entire
initial HTML document is fetched.  This is different from how current browsers
operate, where web sites are progressively rendered as each segment of the
HTML document is received and parsed.

This entails a trade off between rendering delay (since rendering is delayed
until the initial HTML document) and network and device resource use (since,
when a page is classified as \RMA, far fewer resources will be fetched and 
processed).

In this sub-section, we evaluate the rendering delay caused by our classifier,
under several representative network conditions. The rendering delay is equal to
the time to fetch the entire initial HTML document.  We find that the
rendering delay imposed is small, especially compared to the dramatic
performance improvements delivered when a page is \RMA (discussed in more detail
in Section~\ref{sec:mapper}).

\point{Classification Time}
We evaluated the rendering delay imposed by our classifier by measuring the time
taken to fetch the initial HTML for a page, under different network conditions,
and compared it against the time taken for document classification.

\begin{figure}[t]
  \centering
  \includegraphics[width=0.5\textwidth]{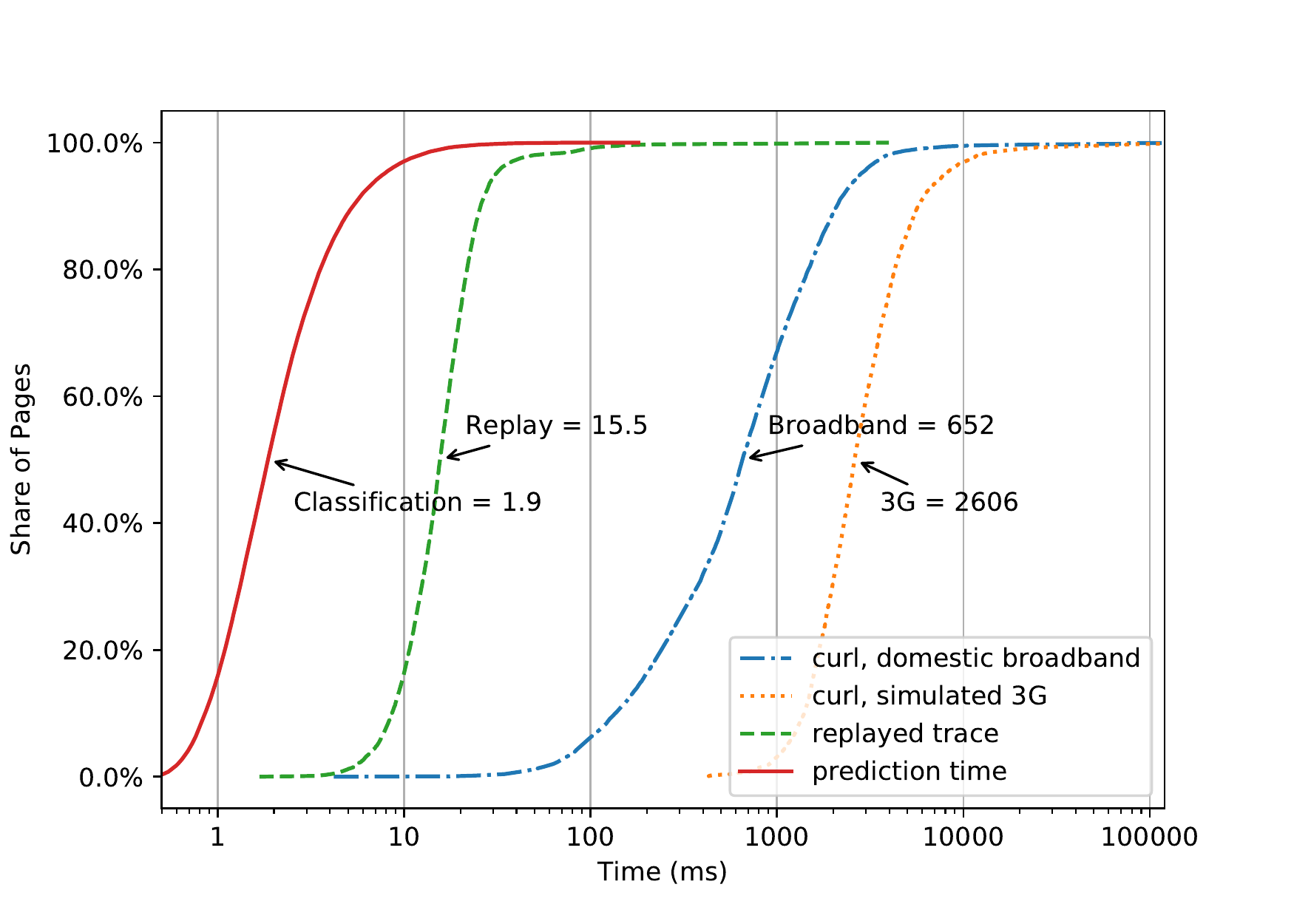}
  \caption{Time to fetch initial HTML document.}
  \label{fig:classifier:time}
\end{figure}

First, we determined how long our classifier took to determine if a parsed HTML
document was \RMA.  We did so by parsing each HTML string with \textit{myhtml},
a fast, open source, \texttt{C++} HTML parser~\cite{myhtml}.  We then measured the execution
time taken to extract the relevant features from the document, and to return the
predicted label.  Our classifier took \ClassifierMeanTime~ms on average and
\ClassifierMedianTime~ms in the median case.

Next, we measured the fixed, simulation cost time of serving each web page
from a locally hosted web server, which allowed us to account for the fixed
overhead in establishing the network connection, and similar unrelated browser
book keeping operations.  This time was \InitHtmlTraceMeanTime~ms on average,
and \InitHtmlTraceMedianTime~ms median.

Finally, we selected two network environments to represent different
network conditions and device capabilities web users are likely to encounter:
a fast, domestic broadband link, with 50~Mbps uplink/downlink
bandwidth and 2~ms latency as indicated by a popular network speed testing
utility\footnote{\url{speedtest.net} - web service that provides analysis of
Internet access performance metrics, such as connection data rate and
latency}, and a simulated 3G network, created using the operating
system's \textit{Network Link Conditioner}\footnote{Network Link Conditioner
is a tool released by Apple with hardware IO Tools for XCode developer tools
to simulate different connection bandwidth, latency and packet loss rates}. We
use a default 3G preset with 780~kbps downlink, 330~kbps uplink, 100~ms packet
delay in either direction and no additional packet loss. Downloading the
documents on such connection took \InitHtmlCurlMeanTime~ms /
\InitHtmlCurlMedianTime~ms (average/median) and \InitHtmlCurlMobileMeanTime~ms
/ \InitHtmlCurlMobileMedianTime ~ms for the two cases respectively.

Figure~\ref{fig:classifier:time} summarizes the results of those measurements.
Overall, the approximately \ClassifierMeanTime~ms taken for an average
document classification is a tiny cost compared to just the initial HTML
download on reasonably fast connections. It could potentially be further
optimized by classifying earlier, i.e. when only a chunk of the initial
document is available. Initial tests show promising results, however this adds
significant complexity to patching the rendering pipeline and we leave it for
future work.

\subsection{Applicability to the Web}
\input{applicability}

While subsequent sections will demonstrate
the significant performance and privacy improvements provided by \ToolName,
these improvements are only available on a certain type of web document,
those that have \RMA subsets.  The performance improvements possible through
\ToolName are therefore bounded by the amount of websites users visit that
are \RMA.

In this subsection we determine how much of the web is amenable to \ToolName,
by applying our classifier to a sampling of websites, representing
different common browsing scenarios.  Doing so allows us to estimate the
benefits \ToolName can deliver, and its relevance to the web.  As presented
in Table~\ref{table:classifier:applicability}, we find that a significant
number of visited URLs are are \RMA, suggesting that \ToolName can deliver
significant privacy and performance improvements to users.  This subsection
continues by describing how we selected URLs in each browsing scenario.

\point{Websites by popularity}
We first estimated how many pages hosted on popular and unpopular domains 
are \RMA.  To do so, we first created two sets of domains, a popular set,
consisting of the most popular 5,000 domains, as determined by Alexa, and
an unpopular set, comprising a random sample of pages ranked 5,001--100,000.

For each domain, we conducted a breath three, depth three crawl.  We first
visited the landing page for the domain, and recorded all linked to pages with
the same TLD+1 domain.  Then we selected up to three urls to from this set, and
repeated the above process another time, giving a maximum of 13 URLs per domain,
and a total data set of \MapperEvalFetchedPages pages. The crawl was conducted
from AWS IP addresses on \MapperEvalDatasetDate.

\point{Social network shared content}
We next estimated how much content linked to from \OSNs is \RMA, to simulate a
user that spends most of their browsing time on popular \OSNs, and generally
only browses away to view shared content.  We gathered URLs shared from Reddit
and Twitter.  We gathered links shared on Reddit by using RedditList
~\cite{redditlist} to obtain top 125 subreddits ranked based on their number of
subscribers. We then visited the 25 posts of each popular subreddit and
extracted any shared URls. For Twitter, we extracted shared links from  the top
10 worldwide Twitter  trends by crawling and extracting shared links from
their Tweets.

\point{RSS / feed readers}
Finally, we estimated how much contend shared from RSS feeds is \RMA, to
simulate a user who finds content mainly through an RSS (or similar)
aggregation service.  We build a list of RSS-shared content by crawling
the Alexa 1K, identifying websites that included RSS feeds, and fetching the
five most recent pages of content in each RSS feed.

\subsection{Conclusion}
In this section we have described how \ToolName determines whether a page
should be rendered in \RM, based on its initial HTML.  We find that \ToolName
outperforms the \StepOne capabilities of existing, deployed \RM tools.  We
also find that the overhead imposed by our classification strategy is small
and acceptable in most cases, and dwarfed by the performance improvements
delivered by \ToolName, for cases when a page is judged \RMA.

%% file: classifier-features.tex
\begin{table}[ht]
    \centering
    \begin{tabular}{@{}lll@{}}
        \toprule
        Number & Name & Description \\ \midrule
            1   & \texttt{p}  & number of \texttt{<p>} \\
            2   & \texttt{ul}&  number of \texttt{<ul>} \\
            3   & \texttt{ol} &  number of \texttt{<ol>} \\
            4   & \texttt{dl}& number of \texttt{<dl>} \\
            5   & \texttt{div}& number of \texttt{<div>} \\
            6   & \texttt{pre}&  number of \texttt{<pre>} \\
            7   & \texttt{table}& number of \texttt{<table>} \\
            8   & \texttt{select}& number of \texttt{<select>} \\
            9   & \texttt{article}& number of \texttt{<article>} \\
            10  & \texttt{section}& number of \texttt{<section>} \\
            11  & \texttt{blockquote} &  number of \texttt{<blockquote>} \\
            12  & \texttt{a}& number of \texttt{<a>} \\
            13  & \texttt{images}& number of \texttt{<img>} \\
            14  & \texttt{scripts}& number of \texttt{<script>} \\
            15  & \texttt{text\_blocks}& \parbox[t]{5cm}{number of text blocks with more than 400 characters (excluding spaces) and are wrapped between 1-11 HTML tags} \\
            16  & \texttt{words} & number of words in \texttt{text\_blocks} \\
            17  & \texttt{url\_depth} & number of subdirectories in URL path \\
            18  & \texttt{amphtml} & 0/1: supports Google AMP \\
            19  & \texttt{fb\_pages} & 0/1: has Facebook channel ID \\
            20  & \texttt{og\_article} & 0/1: has Open Graph article type \\
            21  & \texttt{schema\_org} & 0/1: has schema.org Article/NewsArticle \\ \bottomrule
    \end{tabular}
    \caption{List of all features the classifier uses to predict readability.}
    \label{table:classifier-features}
\end{table}

%% file: classifier-accuracy-dataset.tex
\begin{table}[!t]
\footnotesize
\centering
\setlength{\tabcolsep}{14pt}
\begin{tabular}{+l^r^r}
    \toprule
    	\rowstyle{\bfseries}%
        Data set      &   Number of pages                 & \% Readable \\
    \midrule
        Article pages   &   \ClassifierTrainingArticleNum   & \ClassifierTrainingArticlePct \\
        Landing pages   &   \ClassifierTrainingLandingNum   & \ClassifierTrainingLandingPct \\
        Random pages    &   \ClassifierTrainingRandNum      & \ClassifierTrainingRandPct \\
    \midrule
    	\rowstyle{\bfseries}%
        Total           &   \ClassifierTrainingTotalNum     & \ClassifierTrainingTotalPct \\
    \bottomrule
\end{tabular}
\caption{Description of data set used for evaluating and training ``readability'' classifiers.}
\label{table:classifier-accuracy-dataset}
\end{table}

%% file: classifier-accuracy.tex
\begin{table}[!t]
\footnotesize
\centering
\setlength{\tabcolsep}{14pt}
\begin{tabular}{+l^r^r}
    \toprule
    	\rowstyle{\bfseries}%
        Classifier      &   Precision                           & Recall \\
    \midrule
        ReadabilityJS   &   \ClassifierReadabilityJSPrecision   & \ClassifierReadabilityJSRecall \\
        DOM Distiller   &   \ClassifierDomDistillerPrecision    & \ClassifierDomDistillerRecall \\
        \rowstyle{\bfseries}%
        \MoeClassifier  &   \ClassifierMoePrecision             & \ClassifierMoeRecall \\
    \bottomrule
\end{tabular}
\caption{Accuracy measurements for three classifiers attempting to replicate the
         manual labels described in Table~\ref{table:classifier-accuracy-dataset}.}
\label{table:classifier-accuracy}
\end{table}

%% file: applicability.tex
\begin{table}[!t]
\footnotesize
\setlength{\tabcolsep}{6pt}
\centering
\begin{tabular}{+l^r^r^r}
  \toprule
    \rowstyle{\bfseries}%
    Measurement         & \# measured  & \# readable & \% readable \\
  \midrule
    Popular pages       & \ApplicabilityAlexaFiveK
                        & \ApplicabilityAlexaFiveKReadable
                        & \ApplicabilityAlexaFiveKReadablePct \\
    Unpopular pages     & \ApplicabilityAlexaUnpopular
                        & \ApplicabilityAlexaUnpopularReadable
                        & \ApplicabilityAlexaUnpopularReadablePct \\
    \midrule
    \rowstyle{\bfseries}%
    Total: Random crawl & \ApplicabilityNumUniquePages
                        & \ApplicabilityNumUniquePagesReadable
                        & \ApplicabilityNumUniquePagesReadablePct \\
    % Estimated page views &  \ApplicabilityNumPageViews      & \ApplicabilityNumPageViewsReadable    & \ApplicabilityNumPageViewsReadablePct \\
    \midrule
    Reddit linked       & \ApplicabilityOSNPagesReddit
                        & \ApplicabilityOSNPagesRedditReadable
                        & \ApplicabilityOSNPagesRedditReadablePct \\
    Twitter linked      & \ApplicabilityOSNPagesTwitter
                        & \ApplicabilityOSNPagesTwitterReadable
                        & \ApplicabilityOSNPagesTwitterReadablePct \\
    RSS linked          & \ApplicabilityOSNPagesRSS
                        & \ApplicabilityOSNPagesRSSReadable
                        & \ApplicabilityOSNPagesRSSReadablePct \\
    \midrule
    \rowstyle{\bfseries}%
    Total: OSN          & \ApplicabilityOSNPagesTotal
                        & \ApplicabilityOSNPagesTotalReadable
                        & \ApplicabilityOSNPagesTotalReadablePct \\
  \bottomrule
\end{tabular}
\caption{Measurements of how applicable our readability strategy is under
         common browser use scenarios.}
\label{table:classifier:applicability}
\end{table}

%% file: 4_mapper.tex
\section{Page \uStepTwo}
\label{sec:mapper}

This sections describe how \ToolName generates a \RM presentation of a page,
for pages that have been classified as \RMA.  Our evaluation includes
three possible \RM renderings, each presenting a different trade off between
amount of media included, and performance and privacy improvements.

Generating a \RM presentation of a HTML document can be though of
as translating one tree structure to another: taking the document represented by
the page's initial HTML and generating the document containing a simplified \RM
version. This process of tree mapping is generally known as \StepTwo. We
evaluate \uStepTwo by comparing the performance and privacy improvements of
the three techniques (\RJS, \DD and \BP) described in detail in Section
\ref{sec:background:existing}.

\subsection{Limitations and Bounds}
We note that we did not attempt any evaluation of how users perceive or enjoy
the \RM versions of pages rendered by each considered technique.  We made this
decision for several reasons.

First, two of the techniques (\RJS and \DD) are developed and deployed by large
browser vendors, with millions or billions of users.  We assume that these
large companies have conducted their own evaluation of the usability of their
\RM techniques, and found them satisfactory to users.

Second, performing a large scale evaluation of the subjective usability of
content summarization tools is a costly and time consuming endeavor, beyond
what is possible with typical research budgets.  This again leads us to trust
that the large companies deploying these tools have already conducted such
research and found the results of these tools compelling.

Finally, the third considered \StepTwo technique, Kohlsch{\"u}tter et
al's \BP~\cite{kohlschutter2010boilerplate}, is academic work that includes its
own evaluation, showing that the technique can successfully extract useful
contents from HTML documents. We assume that the authors' evaluation is
comprehensive and sufficient, and that their technique can successfully render
pages in \RM presentations.

\subsection{Evaluation Methodology}
\input{mapper-performance-dataset}

We compared the performance and privacy improvements achieved through
\ToolName's novel application of three \StepTwo techniques: \RJS, \DD and
\BP.  We conducted this evaluation in three stages.

First, we fetched the HTML of each URL in the random crawl data set outlined in
Table~\ref{table:classifier:applicability}, again from an AWS IP. The HTML
considered here is only the initial HTML response, not the state of the document
after script execution. We evaluated whether each of the \MapperEvalFetchedPages
fetched pages were \RMA, by applying the \ToolName classifier to each page.  We
then reduced the data set to the \MapperEvalReadableNum pages
(\MapperEvalReadablePct) were \RMA.

Second, we revisited each URL classified as \RMA to collect a complete version
of the page. To minimize variations in page performance and content during the
testing, we collected the "replay archive" for each page using the "Web Page
Replay" (WPR) performance tool. WPR is used in Chrome's testing framework for
benchmarking purposes and works as a proxy that records network requests or
responds to them instead of letting them through to the source depending on
whether it works in "record" or "replay" mode.

Finally, we applied each of the three \StepTwo techniques to the remaining
\MapperEvalReadableNum HTML documents, and compared the network, resource use,
and privacy characteristics of each transformed page, against the full version
of each page.  We evaluate performance and privacy characteristics of each page
by visiting the URL as replayed from its archive. These findings are described
in detail in the next subsections.

We note that using a replay proxy with a snapshot of content often
underestimates the costs of a page load. Despite taking care to mitigate the
effects of non-determinism by injecting a small script that overrides date
functions to use a fixed date and random number generator functions to use a
fixed seed and produce a predictable sequence of numbers, it cannot account for
all sources of non-determinism. For all requests that the proxy cannot match it
responds with a \texttt{Not Found} response. We notice that it results in a
small number of requests being missed, primarily those responsible for dynamic
ad loading or tracking. It also occasionally interferes with site publisher's
custom resource fetching retry logic, where the same request is retried a number
of times unsuccessfully, before the entire page load times out and the
measurement is omitted.

\subsection{Results: Performance}
\label{subsec:mapper:performance}
\input{mapper-performance}

\begin{figure*}
  \includegraphics[width=\textwidth]{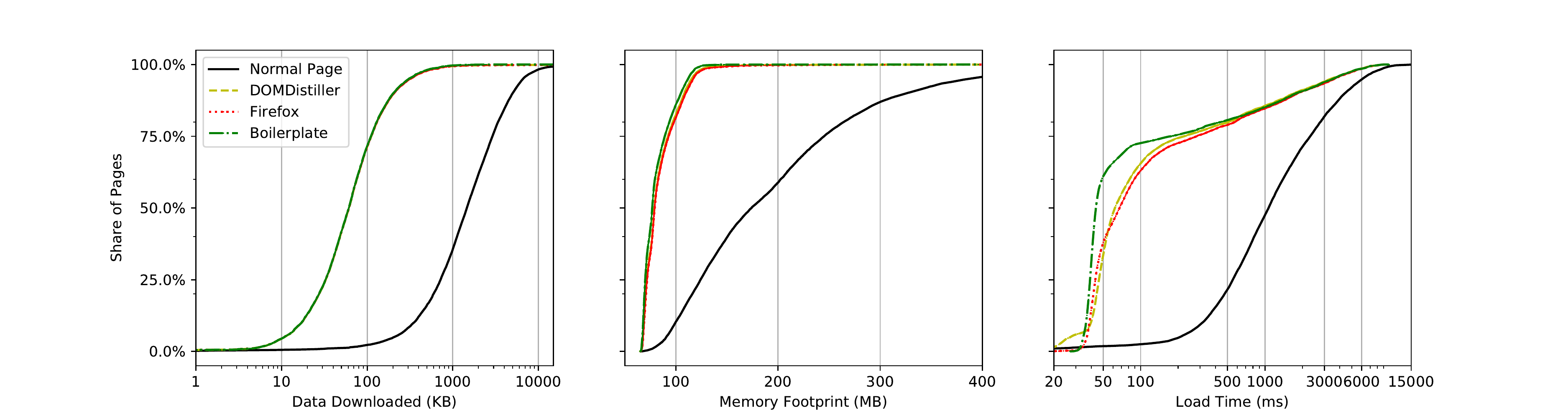}
  \caption{Performance characteristics of the different tree transducer strategies applied, showing the distribution of the key performance metrics.}
  \label{fig:mapper:performance}
\end{figure*}

We measured four performance metrics: number of resources
requested, amount of data fetched, memory used and page load time.
These results summarized in 
Table~\ref{table:mapper:performance} and Figure~\ref{fig:mapper:performance}.

We ran all measurements on AWS \texttt{m5.large} EC2 instances. For performance
measurements, one test was executed at a time, per instance. For each
evaluation, we fetched the page from a previously collected record-replay
archive, with performance tracing enabled. Once the page was loaded and the
performance metrics are recorded, we closed the browser and proxy, and started
the next test.  No further steps were taken to minimize the likelihood of test
VM performance being impacted by interfering workloads on the underlying
hardware.   For all tests we used an unmodified Google Chrome browser, version
70.0.3538.67, rendered in Xvfb\footnote{While Chrome "headless" mode is
available, it effectively employs a different page rendering pipeline with
different load time characteristics and memory footprint.}.

Although profiling has overheads of its own~\cite{Thomas2015}, in particular for
memory use and load times, we used a consistent measurement strategy across all
tests, and therefore expect the impact to also be consistent and minor compared
to relative performance gains.

We measured a page's load time as the difference between
\texttt{navigationStart} and \texttt{loadEventEnd} events~\cite{Wang:12:NT} in
the main frame (i.e. the time until all sub-resources have been downloaded and
the page is fully rendered). Since page content is replayed from a local proxy,
network bandwidth and latency variation impact is minimized and the reported
load time is a very optimistic figure, especially for bigger pages with more
subresources as illustrated in Figure~\ref{fig:classifier:time}. Although
network cost is still non-zero, the number primarily reflects the time taken to
process and render the entire page.

We also recorded the number of resources fetched and the amount of data
downloaded during each test. Note that the amount of data
downloaded for all of the \StepTwo strategies reflects the size of the initial
HTML rather than that of the transformed document, as the transformation happens
on the client and does not result in additional network
traffic. All measured transducers discard the majority of page content (both in
page content like text and markup, but also referenced content like images,
video files, and \JS. Figures~\ref{fig:background:defaultRendering}
and~\ref{fig:background:renderingComparison} provide an example of how
\StepTwo techniques simplify page content.

For memory consumption, we measure the overall memory used by the browser and
its subprocesses. Google Chrome uses a multi-process model, where each tab and
frame may run in a separate process and content of each page also affects what
runs in the main browser process. We note that our testing scenario does not
consider the scenario of multiple pages open simultaneously in the same browsing
session, as some of the resources are reused. The reported number is therefore
that of the entire browser rather than the specific page alone, with some fixed
browser runtime overheads.

Memory snapshots are collected with an explicit trigger after the page load is
complete with  \texttt{disabled-by-default-memory-infra} tracing category
enabled. Despite including a level of fixed browser memory costs, we still see
average memory reduction of up to \MapperMeanMBMemGainMax$\times$ in average or
median cases.

Overall, depending on the chosen transducer, we show:
\begin{itemize}
\item average speedups ranging from \MapperMeanRenderMsGainMin$\times$ to \MapperMeanRenderMsGainMax$\times$
\item average bandwidth savings on the order of \MapperMeanKBFetchedGainMax$\times$
\item number of requests is reduced \MapperMeanRequestsGainMin$\times$ to \MapperMeanRequestsGainMax$\times$
\item average memory reduction of \MapperMeanMBMemGainMax$\times$
\end{itemize}

\subsection{Results: Privacy}
\label{sec:results:privacy}

\input{mapper-privacy}

\ToolName achieves dramatic privacy improvements, because it applies the
\StepTwo step \textit{before} rendering the document, and thus before any
requests to third parties has been initiated.  The privacy improvements gained
by \ToolName are threefold then: a reduction in third parties communicated with,
a reduction in script executed (an often necessary, though not sufficient, part
of fingerprinting online), and a complete elimination ad and tracking related
requests (as labeled by \EL and \EP).  This last measure is particularly 
important, since \textit{92.8\%} of the \MapperEvalDatasetNumPages \RMA pages in
our data set loaded resources labeled as advertising or tracking related by \EL
and \EP~\cite{easylist,github-easylist}.

This subsection proceeds by both describing how we measured the privacy improvements
provided by \ToolName, and the results of that measurement.  These findings are
presented in Table~\ref{table:mapper-privacy}.

We measured the privacy gains provided by \ToolName by first generating \RM
versions of each of the \MapperEvalDatasetNumPages \RMA URLs in our dataset,
and counting the number of third parties, script resources, and ad and tracking
resources in each generated \RM page.  We determined the number of ad and
tracking resources by applying \EL and \EP with an open-source ad-block Node
library \cite{braveadblock} to each resource URL included in the page.  We
then compared these measurements to the number of third-parties, script units,
and ad and tracking resource requests made in the typical, non-\RM rendering
of each URL.

We found that all three of the evaluated \StepTwo techniques dramatically
reduced the number of third parties communicated with, and removed all script
execution and ad and tracking resource requests from the page. Put differently
\ToolName is able to achieve privacy improvements at least as good, and almost
certainly exceeding existing ad and tracking blockers, on \RMA pages. This claim
is based on the observation that ad and tracking blockers do not achieve the
same dramatic reduction in third party communication and script execution as
\ToolName achieves).

%% file: mapper-performance-dataset.tex
\begin{table}[!t]
\footnotesize
\setlength{\tabcolsep}{18pt}
\centering
\begin{tabular}{+l^r}
    \toprule
        \rowstyle{\bfseries}%
        Measurement                 &  Value \\
    \midrule
        Measurement date            &  \MapperEvalDatasetDate \\
        \# crawled domains          &  \MapperEvalFetchedDomains \\
        \# crawled pages            &  \MapperEvalFetchedPages \\
    % \midrule
        \# domains with \RMA pages  &  \MapperEvalDatasetNumDomains \\
        \# \RMA pages               &  \MapperEvalDatasetNumPages \\
        \rowstyle{\bfseries}%
        \% \RMA pages               &  \MapperEvalReadablePct \\
    \bottomrule
\end{tabular}
\caption{Description of data set used for evaluating the performance
         implications of different content extraction strategies.}
\label{table:mapper-performance-dataset}
\end{table}

%% file: mapper-performance.tex
{\footnotesize
\begin{table}[!t]
\setlength{\tabcolsep}{5pt}
\centering
\begin{tabular}{+l^r^r^r^r^r^r^r^r}
    \toprule
        \rowstyle{\bfseries}%
        Transducer      &   \multicolumn{2}{^c}{Resources}
                        &   \multicolumn{2}{^c}{Data}
                        &   \multicolumn{2}{^c}{Memory}
                        &   \multicolumn{2}{^c}{Load Time} \\
                        &   \multicolumn{2}{^c}{(\#)}
                        &   \multicolumn{2}{^c}{(KB)}
                        &   \multicolumn{2}{^c}{(MB)}
                        &   \multicolumn{2}{^c}{(ms)} \\
        \rowstyle{\bfseries}%
       -                &   A & M
                        &   A & M
                        &   A & M
                        &   A & M \\
    \midrule
        Default         &  \MapperDefaultMeanRequests & \MapperDefaultMedianRequests
                        &  \MapperDefaultMeanKBFetched & \MapperDefaultMedianKBFetched
                        &  \MapperDefaultMeanMBMem & \MapperDefaultMedianMBMem
                        &  \MapperDefaultMeanRenderMs & \MapperDefaultMedianRenderMs \\
        ReadabilityJS   &  \MapperReadabilityJSMeanRequests & \MapperReadabilityJSMedianRequests
                        &  \MapperReadabilityJSMeanKBFetched & \MapperReadabilityJSMedianKBFetched
                        &  \MapperReadabilityJSMeanMBMem & \MapperReadabilityJSMedianMBMem
                        &  \MapperReadabilityJSMeanRenderMs & \MapperReadabilityJSMedianRenderMs \\
        Dom Distiller   &  \MapperDomDistillerMeanRequests & \MapperDomDistillerMedianRequests
                        &  \MapperDomDistillerMeanKBFetched & \MapperDomDistillerMedianKBFetched
                        &  \MapperDomDistillerMeanMBMem & \MapperDomDistillerMedianMBMem
                        &  \MapperDomDistillerMeanRenderMs & \MapperDomDistillerMedianRenderMs \\
        BoilerPipe      &  \MapperBoilerPlateMeanRequests & \MapperBoilerPlateMedianRequests
                        &  \MapperBoilerPlateMeanKBFetched & \MapperBoilerPlateMedianKBFetched
                        &  \MapperBoilerPlateMeanMBMem & \MapperBoilerPlateMedianMBMem
                        &  \MapperBoilerPlateMeanRenderMs & \MapperBoilerPlateMedianRenderMs \\
    \midrule
        \rowstyle{\bfseries}%
        Gain ($\times$) \\
    \midrule
        ReadabilityJS   &  \MapperReadabilityJSMeanRequestsGain & \MapperReadabilityJSMedianRequestsGain
                        &  \MapperReadabilityJSMeanKBFetchedGain & \MapperReadabilityJSMedianKBFetchedGain
                        &  \MapperReadabilityJSMeanMBMemGain & \MapperReadabilityJSMedianMBMemGain
                        &  \MapperReadabilityJSMeanRenderMsGain & \MapperReadabilityJSMedianRenderMsGain \\
        Dom Distiller   &  \MapperDomDistillerMeanRequestsGain & \MapperDomDistillerMedianRequestsGain
                        &  \MapperDomDistillerMeanKBFetchedGain & \MapperDomDistillerMedianKBFetchedGain
                        &  \MapperDomDistillerMeanMBMemGain & \MapperDomDistillerMedianMBMemGain
                        &  \MapperDomDistillerMeanRenderMsGain & \MapperDomDistillerMedianRenderMsGain \\
        BoilerPipe      &  \MapperBoilerPlateMeanRequestsGain & \MapperBoilerPlateMedianRequestsGain
                        &  \MapperBoilerPlateMeanKBFetchedGain & \MapperBoilerPlateMedianKBFetchedGain
                        &  \MapperBoilerPlateMeanMBMemGain & \MapperBoilerPlateMedianMBMemGain
                        &  \MapperBoilerPlateMeanRenderMsGain & \MapperBoilerPlateMedianRenderMsGain \\
    \bottomrule
\end{tabular}
\caption{Performance comparisons of three popular readability tree transducer
         strategies, as applied to the data set described in Table~\ref{table:mapper-performance-dataset}.
         Values are given as \underline{A}verage, \underline{M}edians. Gain multiplier (\underline{$\times$}) is calculated for each page load and \underline{A}verage and \underline{M}edian values are reported.}
\label{table:mapper:performance}
\end{table}}

%% file: mapper-privacy.tex
\begin{table}[!t]
\footnotesize
\setlength{\tabcolsep}{8pt}
\centering
\begin{tabular}{+l^r^r^r^r^r^r}
    \toprule
        \rowstyle{\bfseries}%
        Transducer      &   \multicolumn{2}{^c}{\# third-party}
                        % &   \multicolumn{3}{c}{Reduction}
                        &   \multicolumn{2}{^c}{\# scripts}
                        % &   \multicolumn{3}{c}{Reduction}
                        &   \multicolumn{2}{^c}{Ads \& Trackers} \\
        \rowstyle{\bfseries}%
                        % &   \multicolumn{3}{c}{Reduction} \\
                        &   Avg & Med
                        % &   Avg & Med & 90\%
                        &   Avg & Med
                        % &   Avg & Med & 90\%
                        % &   Avg & Med & 90\%
                        &   Avg & Med \\
    \midrule
        Default         &  \MapperDefaultMeanTP & \MapperDefaultMedianTP
                        % & - & - & -
                        &  \MapperDefaultMeanScripts & \MapperDefaultMedianScripts
                        % & - & -   & -                   
                        &  \MapperDefaultMeanLabeled & \MapperDefaultMedianLabeled \\
                        % & - & - & - \\
        ReadabilityJS   &  \MapperReadabilityJSMeanTP & \MapperReadabilityJSMedianTP
                        % &  \MapperReadabilityJSMeanTPSaved & \MapperReadabilityJSMedianTPSaved & \MapperReadabilityJSNinetyTPSaved
                        &  \MapperReadabilityJSMeanScripts & \MapperReadabilityJSMedianScripts
                        % &  \MapperReadabilityJSMeanScriptsSaved & \MapperReadabilityJSMedianScriptsSaved & \MapperReadabilityJSNinetyScriptsSaved
                        &  \MapperReadabilityJSMeanLabeled & \MapperReadabilityJSMedianLabeled \\
                        % &  \MapperReadabilityJSMeanLabeledSaved & \MapperReadabilityJSMedianLabeledSaved & \MapperReadabilityJSNinetyLabeledSaved \\
        Dom Distiller   &  \MapperDomDistillerMeanTP & \MapperDomDistillerMedianTP
                        % &  \MapperDomDistillerMeanTPSaved & \MapperDomDistillerMedianTPSaved & \MapperDomDistillerNinetyTPSaved
                        &  \MapperDomDistillerMeanScripts & \MapperDomDistillerMedianScripts
                        % &  \MapperDomDistillerMeanScriptsSaved & \MapperDomDistillerMedianScriptsSaved & \MapperDomDistillerNinetyScriptsSaved
                        &  \MapperDomDistillerMeanLabeled & \MapperDomDistillerMedianLabeled \\
                        % &  \MapperDomDistillerMeanLabeledSaved & \MapperDomDistillerMedianLabeledSaved & \MapperDomDistillerNinetyLabeledSaved \\
        BoilerPipe      &  \MapperBoilerPlateMeanTP & \MapperBoilerPlateMedianTP
                        % &  \MapperBoilerPlateMeanTPSaved & \MapperBoilerPlateMedianTPSaved & \MapperBoilerPlateNinetyTPSaved
                        &  \MapperBoilerPlateMeanScripts & \MapperBoilerPlateMedianScripts
                        % &  \MapperBoilerPlateMeanScriptsSaved & \MapperBoilerPlateMedianScriptsSaved & \MapperBoilerPlateNinetyScriptsSaved
                        &  \MapperBoilerPlateMeanLabeled & \MapperBoilerPlateMedianLabeled \\
                        % &  \MapperBoilerPlateMeanLabeledSaved & \MapperBoilerPlateMedianLabeledSaved & \MapperBoilerPlateNinetyLabeledSaved \\
    \bottomrule
\end{tabular}
\caption{Comparisons of the privacy implications of three popular readability tree transducer
         strategies, as applied to the data set described in Table~\ref{table:mapper-performance-dataset}.
         Values are given as Average and Median values.}
\label{table:mapper-privacy}
\end{table}

%% file: 6_discussion.tex
\section{Discussion and Future Work}
\label{sec:discussion}

\subsection{\uuRM as a Content Blocker}
Most existing \RM tools function to improve the presentation of page content
for readers, by removing distracting content and reformatting text for the
browser user's benefit.  While the popularity of existing \RMs suggest that
this is a beneficial use case, the findings in this work suggest an additional
use case for \RMs, blocking advertising and tracking related content.

As discussed in Section~\ref{sec:results:privacy}, \ToolName prevents all
ad and tracking related content from being fetched and rendered, as identified
by \EL and \EP (Table~\ref{table:mapper-privacy}).  \ToolName also loads between
\MapperMeanRequestsGainMin and \MapperMeanRequestsGainMax~times fewer resources
than typical page rendering and \RMs (Table~\ref{table:mapper:performance}), a 
non-trivial number of which are are likely also ad and tracking related.

\ToolName differs fundamentally from existing content blocking strategies.
Existing popular tools, like uBlock Origin\cite{ublockorigin} and
AdBlock Plus\cite{adblockplus}, aim to identify \textit{malicious} or undesirable content,
and prevent it from being loaded or displayed; all unlabeled content is treated
as desirable and loaded as normal.  \ToolName, and (at last conceptually) \RMs 
in general, take the opposite approach.  \RMs try to identify \textit{desirable}
content, and treat all other page content as undesirable, or, at least, unneeded.

Our results suggest that the \RM technique can achieve ad and tracking blocking
quality \textit{at least} as well as existing content blocking tools, but with
dramatic  performance improvements.  We expect that \ToolName actually outperforms
content blocking tools (since content blockers suffer from false-negative problems,
for a variety of reasons), but lack a ground truth basis to evaluate this claim
further.  We suggest evaluating the content blocking capabilities of \RM-like tools
as a compelling area for future work.

\subsection{Comparison to Alternatives}
\ToolName exists among other systems that aim to improve the user experience of viewing
content on the web. While a full evaluation of these systems is beyond the scope of
this work (mainly because the compared systems have different goals and place
different restrictions on users), we note them here for completeness.

\point{AMP} Accelerated Mobile Pages (AMP)\cite{ampproject} is a system
developed by Google that improves website performance, in a number of ways.
Website authors opt-in to the AMP system by limiting their content to a subset
of HTML, \JS and CSS functionality, which allows for optimized loading and
execution.  AMP pages are also served from Google's servers, which provide
network level improvements.  AMP differs from \ToolName and other \RM systems in
that users only achieve performance improvements when site authors design their
pages for AMP; AMP offers no improvement on existing, traditional websites.

\point{Server-Assisted Rendering} Other browser vendors attempt to improve the user
experience by moving page, loading, rendering execution from the client to a server.
The client then fetches a rendered version of the page from the server (generally
either rendered HTML or as a bitmap).  The most popular such system is likely
Amazon Silk\cite{amazonsilk}.  While there are significant performance upsides with
this thin-client technique, they come with significant downsides too.  First, user
privacy is harmed, since the rendering-server must manage and observe all client secrets
when interacting with the destination server on the client's behalf.  Additionally,
while the server may be able to improve the loading and rendering of the page,
its limited in the kinds of performance improvements it can achieve.  Server assisted
rendering does not provide any of the presentation simplification or content blocking
benefits provided by \ToolName.)

\subsection{\ToolName Deployment Strategies}
\point{Always On} \ToolName as described in this work is designed to be ``always on'',
attempting to provide a \RMA presentation of every paged fetched.  This decision
entails several tradeoffs.  It maximizes the amount of privacy and performance
improvements provided, but entails an overhead while loading each page (Figure 
\ref{fig:classifier:time}).  This overhead may not be worthwhile in some browsing
patterns (e.g. users who use the browser mostly to interact with application like sites).

Additionally, there may be times when users want to maintain a page's interactive
functionality (e.g. \JS), even when \ToolName has determined that the page is \RMA.
Making sure that the user has the ability to disable \ToolName would be important
in such cases.  The system described in this work does not preclude such an option,
but only imagines changing the default page loading behavior\footnote{Current
browsers and \RMs load all pages in the standard manner, and allow the users to
enable a \RM presentation, while \ToolName would load pages in the optimized
\RM presentation by default, when possible, and allow users to enable the standard
loading behavior.}.

\point{\uuStepTwo Improvements} Section~\ref{sec:mapper} evaluates three possible
\StepTwo techniques, each of which can provide a \RM presentation with different
performance and privacy improvements.  These three evaluated techniques are
adapted from existing tools and research. Users of \ToolName could select which
\StepTwo technique best suited their needs.

However, we expect that ML and similar techniques could be applied to the \StepTwo
problem, to provide a \RM presentation that exceeds existing techniques.  An improved
\StepTwo algorithm would achieve equal or greater performance and privacy
improvements, while doing a better job of maintaining the meaning and information of
the extracted content.  We are currently exploring several options in this area, but
have found the problem large enough to constitute its own unique work.

%% file: 7_related_work.tex
\section{Related Work}
\label{sec:related-work}

\subsection{Content Extraction }

The problem of
removing boilerplate and extracting relevant content from a Web page has been
extensively studied. Since the user's visual perception of a Web
page is vastly different from its HTML structure, previous approaches
primarily focused on the code structure, visual representation and the link
between the two. Lin \etal~\cite{lin2002discovering} proposed a method to detect
content blocks using \texttt{<TABLE>} tags and calculate their entropy to
distinguish the informative blocks from the redundant ones. Laber 
\etal~\cite{laber2009fast} proposed a heuristic method for extracting textual
sections and title from news articles using \texttt{<a>}, \texttt{<p>} and
\texttt{<title>} tags.

Other studies have tried to detect \emph{useful} segments in a web page using
structural and positional information.  Gupta
\etal~\cite{Gupta:2003:DCE:775152.775182} introduced a DOM-based method to
modify and remove irrelevant DOM nodes to extract the main content. Their
approach utilized filters to remove DOM nodes with advertisements, and link and
text ratio thresholds to remove unwanted table cells. While the proposed
rule-based method was simple, it had a poor performance in link rich pages where
the main content contained many links. Weninger \etal~\cite{weninger2010cetr}
introduced a fast algorithm which calculates the HTML tag ratio of each line to
cluster and extract text content. Their algorithm did not perform well on home
pages as well as it suffered from high recall and low precision.
Cai~\etal~\cite{vips-a-vision-based-page-segmentation-algorithm} introduced a
tag-free vision-based page segmentation algorithm to segment a Web page and
extract its Web content structure using the link between the visual layout and
the content. Fan \etal~\cite{fan2011article} introduced Article Clipper, a Web
content extractor that leveraged visual cues in addition to HTML tags to extract
non-textual and textual content and detect multi-page articles. Their approach
underperformed in extracting captions which were links as well as images and
captions that were outside of main content.

Heuristic methods are limited by their lack of adaptability.
Some have proposed learning based methods to overcome this rigidness.
Pasternack and Roth~\cite{Pasternack:2009:EAT:1526709.1526840}
describe a semi-supervised algorithm called Maximum Subsequence Segmentation,
which tokenized HTML into list of tags, words and symbols, and attempts to
classify each block as either "in article text" or "out of article" text.
Kohlsch{\"u}tter
\etal~\cite{kohlschutter2010boilerplate} developed BoilerPipe to classify text elements
using both structural and text features, such as average word length and average
sentence length. Sun \etal~\cite{sun2011dom} proposed Content Extraction via
Text Density (CETD) to extract the text content from pages using a variety of
text density measurements. Their method relied on the observation that the
amount of text in content sections is large, and the text in boilerplate
sections contains more noise and links. Sluban and
Gr{\v{c}}ar~\cite{sluban2013url} introduced an unsupervised and
language-independent method for extracting content from streams of HTML pages,
by detecting commonalities in page structure. While their method outperformed
other open source content extractor algorithms, it suffered from high memory
consumption and poor performance in diverse and small HTML dataset.

Wu \etal~\cite{wu2015automatic} proposed a machine learning model using DOM
tree node features such as position, area, font, text and tag properties to
select and group content related nodes and their children. In their recent
paper, Vogels \etal~\cite{vogels2018web2text} presented an algorithm combining
a hidden markov model in and a convolutional neural networks (CNNs).  Their model first
preprocessed an HTML page into a Collapsed DOM (CDOM) tree where each single
child parent node was merged with its child. CDOM was then segmented into
blocks of main content and boilerplate using sequence labeling of DOM leaves.
The features were then used to train two CNNs, obtain potentials and finally
find the optimal labeling. Their approach outperformed previous studies on the
CleanEval benchmark~\cite{baroni2008cleaneval}.

\subsection{Web Complexity}

While content extraction has attracted much
attention in the scientific literature, fewer studies are conducted to
understand the complexity of websites and its impact on page load time and
user experience.  Gibson \etal~\cite{gibson2005volume} analyzed Web page
template evolution using site-level template detection algorithms and found
that templates, with little raw content value, represented 40-50\% of the data
on the Web and the rate continued to grow at a rate about 6\% per year. In
their large-scale study, Butkiewicz \etal~\cite{butkiewicz2011understanding}
showed that modern websites, regardless of their popularity, were complex and
such complexity could affect user experience. 
Moreover, their analysis demonstrated
that the number of loaded objects and the number of servers could indicate Web
page load time. Their result showed News websites loaded significantly more
objects from more servers.

\subsection{Performance and User Experience}

While complexity of Web pages can
affect page load time,  visual complexity of a Web page can impact user
experience. Harper \etal showed that visual complexity in Web pages, defined
as diversity, density, and positioning of the elements, could increase
cognitive load~\cite{Harper:2009:TDV:1498700.1498704} and even have
detrimental cognitive and emotional impact on users~\cite{TUCH2009703}. In
many websites, online advertisements are the only source of income.
Nonetheless, online ads, especially intrusive ads, have usability 
consequences~\cite{brajnik2010review}. As Pujol 
\etal~\cite{Pujol:2015:AUA:2815675.2815705} observed, 22\% of the most active users
of a major European ISP use Adblock Plus. As a result, providing the main
content in a simple and less cluttered page, such as Reader Mode, not only
decreases the complexity of a page, but also can preserve privacy by limiting
the number of requests for third-party services and 
trackers~\cite{Krishnamurthy:2009:PDW:1526709.1526782,
Englehardt:2016:OTM:2976749.2978313} as well as improve user experience.

%% file: 8_conclusion.tex
\section{Conclusion}
\label{sec:conclusion}

The modern web's progress has led us to the point far beyond Hypertext Markup
for document discovery, to having full-fletched, media-rich experiences and
dynamic applications. With this growth in capability has been a growth
in page ``bloat'', making pages expensive to load, and bringing with it
ubiquitous advertising and tracking. In this work we propose \ToolName as an
approach broadening the applicability of ``reader mode'' browser features to
delivers huge improvements to the end-user browsing experience.

Unique among \RM tools, \ToolName determine if a page is \RMA based only on
the page's initial HTML, before the HTML is parsed and rendered, and before
sub resources are fetched. Or classifier within 2~ms with 91\% accuracy, making it
practical as an always-on part of the rendering pipeline, transforming all suitable pages at load time. We
find that \ToolName is widely applicable, and can deliver is performance and
privacy improvements to at 22\% of pages on popular and unpopular websites, and a larger
proportion of pages linked to from \OSN like Reddit (42\%) and Twitter (31\%).

Because \ToolName makes its modifications before sub-resources are fetched,
\ToolName uses \MapperMeanKBFetchedGainMax$\times$ less network than traditional
page rendering (and current \RM techniques). This results in page load time
improvements, important in a range of scenarios from poor connectivity or
low-end devices, to expensive data connectivity or simply wanting a clean and
simple interaction with primarily textual content. \ToolName also delivers page
loading speedups of \MapperMeanRenderMsGainMin$\times$ -
\MapperMeanRenderMsGainMax$\times$ and average memory reduction of
\MapperMeanMBMemGainMax$\times$, while maintaining a pleasant, \RM style user
experience.

Finally, we find that the \ToolName excels at protecting user's privacy by
removing effectively all currently recognized third-party trackers and ads from
the tested pages. When applied to \MapperEvalReadableNum \RMA webpages, 
\ToolName prevented 100\% of advertising and tracking related resources from 
being fetched (as labeled by \EL and \EP).